# A switchable two-dimensional electron gas based on ferroelectric Ca:SrTiO$_3$

Julien Bréhin[1♣], Felix Trier[1♣], Luis M. Vicente-Arche[1♣], Pierre Hemme[2], Paul Noël[3], Maxen Cosset-Chéneau[3], Jean-Philippe Attané[3], Laurent Vila[3], Anke Sander[1], Yann Gallais[2], Alain Sacuto[2], Brahim Dkhil[4], Vincent Garcia[1], Stéphane Fusil[1], Agnès Barthélémy[1], Maximilien Cazayous[2] and Manuel Bibes[1*]

[1] Unité Mixte de Physique CNRS, Thales, Univ. Paris-Sud, Université Paris-Saclay, 91767 Palaiseau, France

[2] Laboratoire Matériaux et Phénomènes Quantiques (UMR 7162 CNRS), Université Paris Diderot-Paris 7, 75205 Paris Cedex 13, France

[3] Univ. Grenoble Alpes, CEA, CNRS, Grenoble INP, Spintec, 38000 Grenoble, France

[4] Laboratoire Structures, Propriétés et Modélisation des Solides, CentraleSupélec, Université Paris Saclay, CNRS UMR8580, 91190 Gif-Sur-Yvette, France

Two-dimensional electron gases (2DEGs) can form at the surface of oxides and semiconductors or in carefully designed quantum wells and interfaces. Depending on the shape of the confining potential, 2DEGs may experience a finite electric field, which gives rise to relativistic effects such as the Rashba spin-orbit coupling. Although the amplitude of this electric field can be modulated by an external gate voltage, which in turn tunes the 2DEG carrier density, sheet resistance and other related properties, this modulation is volatile. Here, we report the design of a "ferroelectric" 2DEG whose transport properties can be electrostatically switched in a non-volatile way. We generate a 2DEG by depositing a thin Al layer onto a SrTiO$_3$ single crystal in which 1% of Sr is substituted by Ca to make it ferroelectric. Signatures of the ferroelectric phase transition at 25 K are visible in the Raman response and in the temperature dependences of the carrier density and sheet resistance that shows a hysteretic dependence on electric field as a consequence of ferroelectricity. We suggest that this behavior may be extended to other oxide 2DEGs, leading to novel types of ferromagnet-free spintronic architectures.

♣ : these authors contributed equally to this work

* manuel.bibes@cnrs-thales.fr

**I. Introduction**

Strontium titanate SrTiO$_3$ (STO) is possibly the most widely used perovskite oxide and one of the richest in terms of functionalities. In its stoichiometric form its dielectric constant – already very large at room temperature ($\varepsilon \approx 300$) – strongly increases upon cooling, reaching values higher than 20000[1]. This behavior, referred to as quantum paraelectricity[2], is very rare in condensed matter and signals the proximity to a ferroelectric state. Indeed, the introduction of a slight structural disorder through the substitution of just 1% of Sr by Ca[3], or by replacing $^{16}$O by $^{18}$O[4] stabilizes a proper ferroelectric state with a Curie temperature of 25-50 K. Furthermore, although bulk STO is a wide bandgap semiconductor, minute n-type doping (by replacing Sr by La, Ti by Nb, or by introducing oxygen vacancies) induces a transition to a metallic state[5] with very high electron mobility (>10$^4$ cm²/Vs) at low temperature[6,7]. At mK temperatures, n-type STO even becomes superconducting[8], qualifying as the most dilute superconductor[9].

For the last 15 years, STO has also served as a platform to generate oxide two-dimensional electron gases (2DEGs) through the epitaxial growth of a polar perovskite such as LaAlO$_3$ (LAO)[10], the deposition of thin reactive metal films (such as Al)[11,12] or by fracturing in vacuum[13]. Although the precise mechanisms remain debated[14], the formation of this n-type 2DEG often involves oxygen vacancies and is reminiscent of the ease by which bulk STO can be doped n-type. Even though bulk STO is thus well established to possess two instabilities[15] – from a wide bandgap dielectric to a n-type metal or to a ferroelectric – only the former has been exploited in the vast literature on STO 2DEGs. Introducing ferroelectricity in STO 2DEGs would provide a means to achieve a non-volatile electric control of its electronic and spin-orbitronic properties among other possible features. Here, we take advantage of both instabilities of STO simultaneously to design a "ferroelectric" 2DEG.

The interplay between ferroelectricity and transport in oxide 2DEGs has already attracted some attention and heterostructures combining LAO/STO with ferroelectrics such as Pb(Zr$_x$Ti$_{1-x}$)O$_3$ (PZT) have been reported[16–18]. A large modulation of the 2DEG resistance and the local carrier density were found[18] and interpreted as due to ferroelectric field effect causing polarization-dependent band bending and charge accumulation/depletion[16]. While these results are notable for applications in oxide electronics, they concern the use of a 2DEG as a channel in a field-effect transistor with a ferroelectric gate oxide rather than the engineering of a new 2DEG ground state. In parallel, recent work reported metallic and superconducting behavior in slightly doped STO thin films grown on substrates imposing a compressive strain of about -1%[19], known to induce a ferroelectric character[20,21]. These studies are motivated by the intriguing increase of the superconducting critical temperature (T$_C$)[22] believed to result from the interplay of superconductivity with ferroelectricity[23]. However, they did not evidence that the transport properties could be switched by ferroelectricity, and to date there are no reports of 2DEGs based on ferroelectric materials.

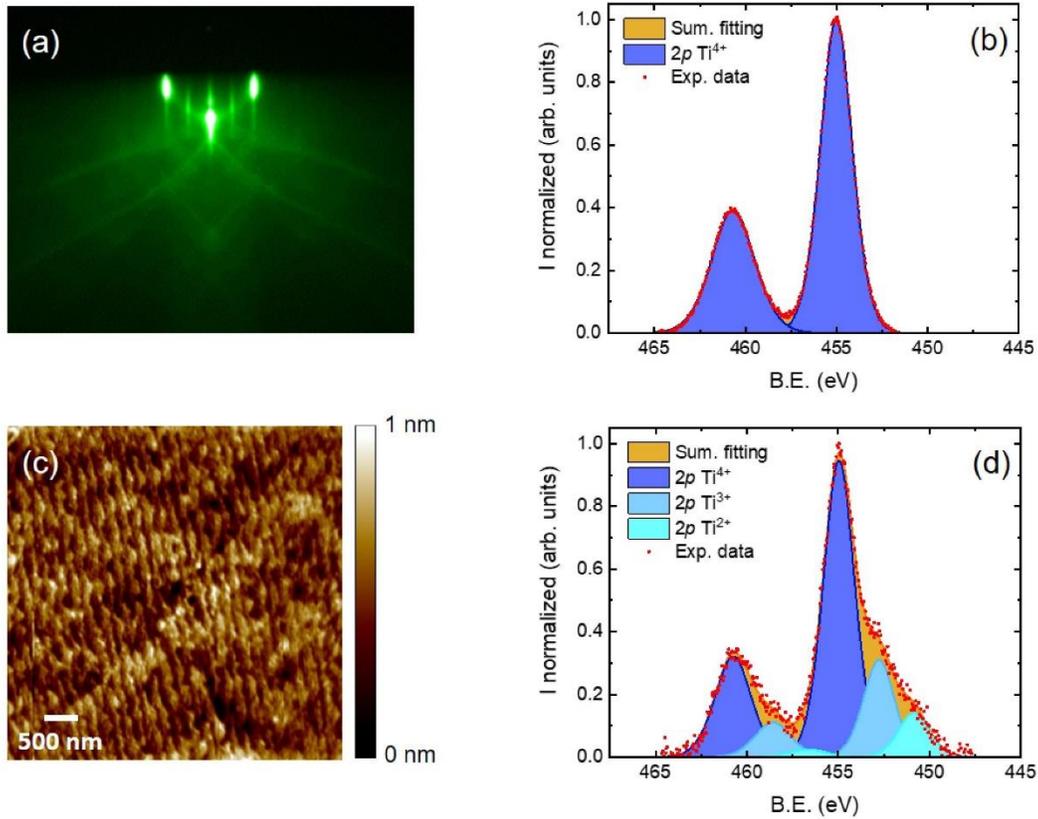

*Figure 1. (a) Reflection high-energy electron diffraction (RHEED) diffraction pattern acquired on the pre-annealed Ca:STO substrate, revealing its high crystal quality. (b) X-ray photo electron spectroscopy (XPS) of the Ti 2p spectra of bare pre-annealed Ca:STO substrate, showing no presence of $Ti^{3+}$. (c) Atomic force microscopy image of an as-grown Al//Ca:STO sample confirms the surface cleanliness. The scale bar corresponds to 500 nm. (d) XPS Ti 2p spectra of the Al//Ca:STO samples, revealing a significant presence of $Ti^{3+}$ amounting to 32.8% relative to the measured $Ti^{4+}$ intensity. Note that all peaks are shifted down by about 3 eV with respect to the tabulated peak positions, presumably due to charging effects.*

## II. Sample preparation

The as-received (001)-oriented $Sr_{0.99}Ca_{0.01}TiO_3$ (Ca:STO) substrates (from SurfaceNet GmbH) were initially pre-annealed for 1 hour at $T$ = 800°C and at an oxygen partial pressure of $P_{O2}$ = 400 mbar in order to remove organic surface contaminants and possible oxygen vacancies. The crystallinity of Ca:STO substrates surface was then confirmed in situ by reflection high-energy electron diffraction (RHEED) imaging, see Figure 1(a).

Following this pre-annealing step, the Ti oxidation state was quantified by in situ X-ray photo-electron spectroscopy (XPS) using a Mg Kα source ($hv$ = 1253.6 eV), showing a pure $Ti^{4+}$ oxidation state, without the detectable presence of $Ti^{3+}$ (see Figure 1(b)). This is consistent with an intrinsically insulating Ca:STO substrate. In the same vacuum cycle, we then deposited a 1.8 nm thick Al film by dc sputtering. XPS Ti 2$p$ spectra after Al deposition shows the clear presence of $Ti^{3+}$ (see Figure 1(d)) with a normalized intensity ratio between the $Ti^{3+}$ and $Ti^{4+}$ peak areas of 32.8% (as analyzed with the CasaXPS software). The magnitude of this ratio is consistent with the presence of a 2DEG at the Al//Ca:STO heterointerface possessing a carrier density in the range of $1 \times 10^{14}$ cm$^{-2}$ at room temperature, similar to the Al//STO case[24,25]. Following the final film deposition, the Al//Ca:STO sample surface was probed with atomic force microscopy and confirmed to be smooth with clearly detectable ∼4 Å lattice terraces (see Figure 1(c)). As a final step, a metal film of Ti(10 nm)/Au(50 nm) was deposited on the back side of the Ca:STO substrate, to be used as a back-gate electrode for the electrostatic tuning of the ferroelectric 2DEG and as a bottom electrode for the polarization measurements.

**III. Physical properties**

To investigate the ferroelectric properties of Al//Ca:STO samples, a triangular waveform was applied at a frequency $f$=100 Hz (using a Radiant Multiferroic system) across the Ca:STO, between the 2DEG and the back electrode, and the current $I$ was measured in real time. Integrating the current with time and normalizing by the sample area yields the polarization. As seen in Figure 2(a), the sample shows clear current peaks in the $I(V)$ curves and corresponding polarization hysteresis loops at low temperature ($T$ = 2 K) consistent with the expected ferroelectric behavior of Ca substituted STO[3]. In order to obtain the transition temperature for ferroelectricity we measured the temperature dependence of the remanent polarization after applying an electric field of ±1.4 V/cm (see Figure 2(b)). As visible in Figure 2(b), the ferroelectric polarization disappears around a temperature $T_C$=25-30 K, consistent with earlier reports[3].

To further substantiate the ferroelectric state in our samples, we performed Raman scattering measurements with a 532 nm laser line from a solid-state laser. The laser spot size was about 100 μm and a small laser power was used to keep the laser heating as low as possible. The spectra were recorded between 10 and 40 K using a Jobin Yvon T64000 triple spectrometer equipped with a liquid-nitrogen-cooled CCD detector. We have been able to detect the Raman signal of Al//Ca:STO at energies as low as 10 cm$^{-1}$. Figure 2(c) shows the Raman spectra at low energy between 10 and 75 cm$^{-1}$ as a function of temperature. Two phonons modes can be observed at 20 and 55 cm$^{-1}$. Consistent with previous reports on ferroelectricity in Ca substituted STO[22], the vibrational mode with a wavenumber of about 20 cm$^{-1}$ in the Raman spectra corresponds to a ferroelectric soft mode (TO1), which moves towards lower energies upon warming. The ferroelectric state is characterized by the activation of this optical phonon modes in the Raman spectrum due to the loss of inversion symmetry. From the relative peak intensity[26], we extract once again a transition temperature of ~25 K (see Figure 2(d)).

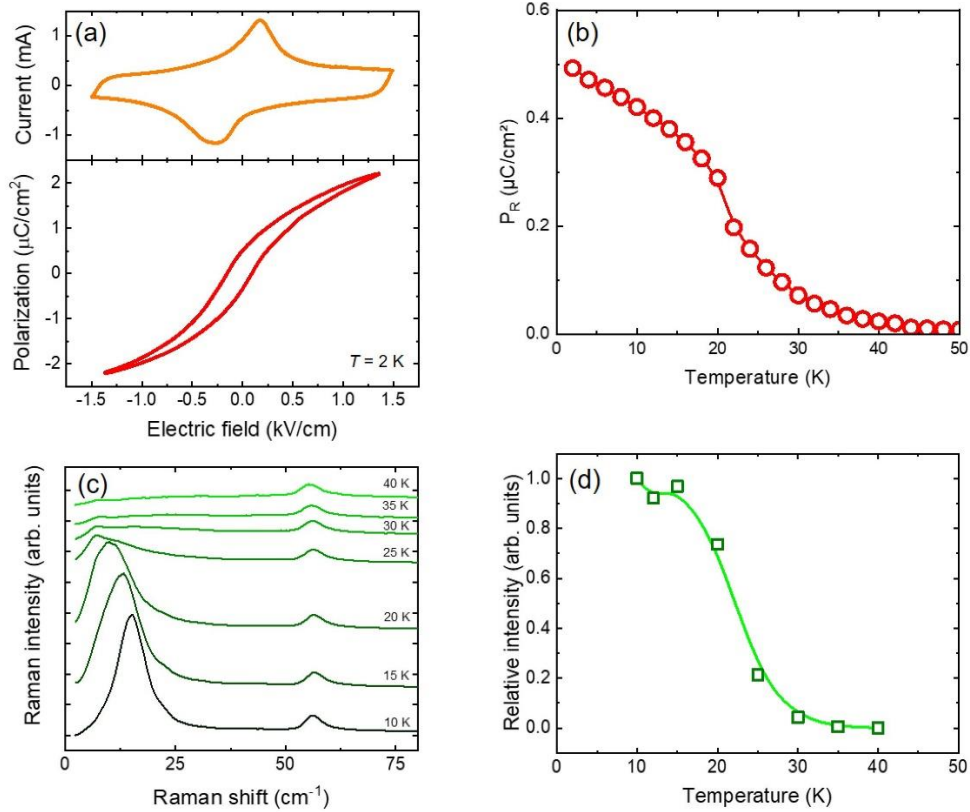

*Figure 2.* (a) Measured current (top panel) and polarization (bottom panel) loops as a function of the electric field for the Al//Ca:STO sample at 2 K are consistent with the existence of a ferroelectric state (f=100 Hz). (b) Temperature dependence of the remanent polarization $P_R$ after the application of a maximum electric field of ±1.4 V/cm, showing a critical transition temperature of ~25 K. (c) Raman spectra of Al//Ca:STO obtained for temperature between 10 K and 40 K, displaying a clear phonon mode present at a shift of 20 cm$^{-1}$ at 10 K. (d) The temperature dependence of the relative intensity of the mode at 20 cm$^{-1}$ in the Raman spectra confirms a critical transition temperature of ≈25 K.

Low temperature electrical transport measurements were performed on the Al//Ca:STO samples bonded by Al wires in the van der Pauw configuration using a standard AC lock-in technique ($I_{AC}$ = 200 nA, $f_{AC}$ = 77.03 Hz) in a Quantum Design Dynacool cryostat. These measurements were carried out at a temperature of 2 K with magnetic fields between -9 T and 9 T for the Hall resistance study. Prior to any actual measurements as a function of the back-gate voltage, the samples were subjected to a so-called forming step[27] at 2 K where the back-gate voltage was cycled several times (>2) between the gate voltage extremes of the particular gate-voltage interval. This ensures that no irreversible changes different from the switching of the ferroelectric state in Ca:STO occurs in the upon application of the back-gate voltage in the actual experiment. Note that this low temperature forming step was repeated following all occasions the sample was brought above 105 K, which corresponds to the ferroelastic transition temperature of STO. Moreover, at each new cooldown, the samples were always cooled with the back-gate electrostatically grounded.

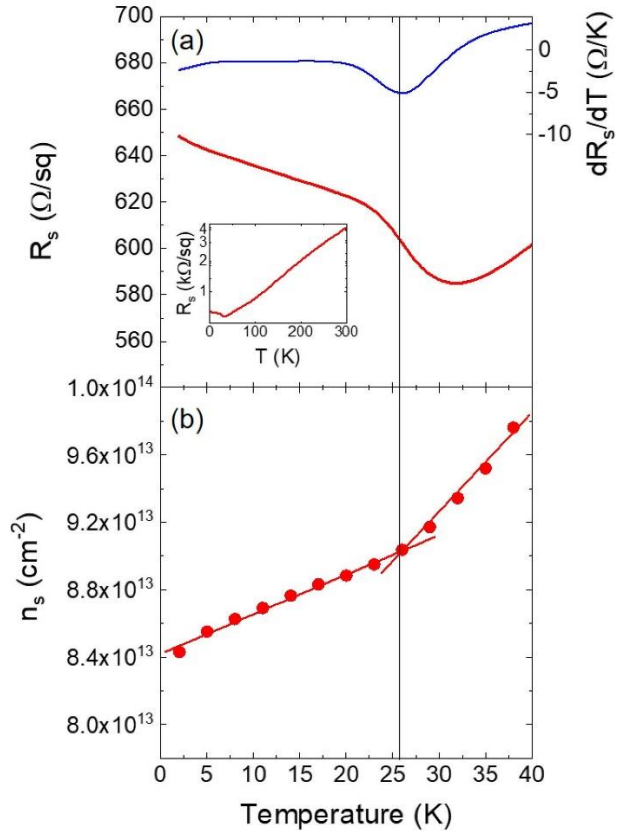

*Figure 3.* (a) Temperature dependence of the sheet resistance, $R_s$, of the Al//Ca:STO sample between 2-40 K (red curve, left y-axis) as well as the sheet resistance derivative with respect to temperature (bmue curve, right y-axis). Both curves point towards a transition temperature around 25 K. The inset shows the $R_s(T)$ between 2-300 K which is representative of usual 2DEG metallic behavior (above the ferroelectric Curie temperature). (b) The Hall carrier density, $n_s$, extracted between 2 and 40 K confirms the presence of two different electronic regimes with a transition temperature of ~25 K. The solid lines are guides to the eye.

The transport measurements in the virgin, ungated state confirmed the globally metallic behavior of the 2DEG with a reduction of the sheet resistance, $R_s$, from 4000 Ω/sq to 650 Ω/sq when cooling from 300 K to 2 K, respectively (see Figure 3(a)). On closer inspection, the temperature dependence of $R_s$ reveals a resistance minimum around 20-40 K not present in usual STO-based 2DEGs[28]. Such a kink in the $R_s$ vs $T$ was interpreted as a fingerprint of the emergence of ferroelectricity in n-type bulk Ca:STO[22] and n-type strained STO thin films[29]. The derivative of the sheet resistance with respect to the temperature reveals a transition temperature around 25 K. By extracting the Hall carrier density as a function of temperature, this transition is likewise found to separate the carrier density evolution into two different regimes with correspondingly different slopes as a function of temperature (see Figure 3(b)).

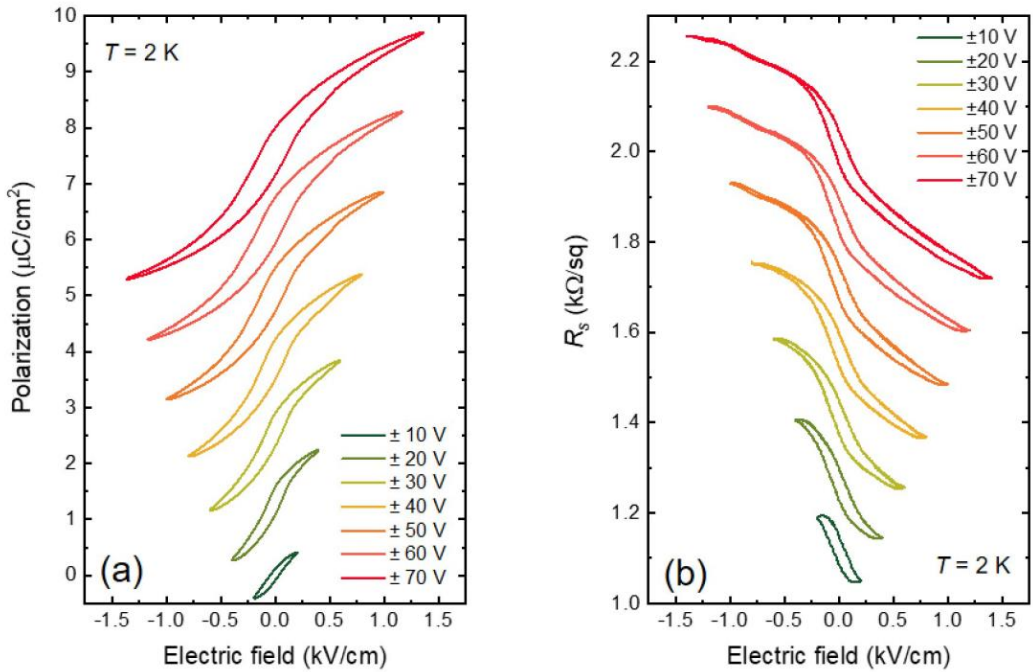

*Figure 4.* (a) Polarization loops at 2 K measured in the ferroelectric state for different increasing maximum values of the back-gate voltage, $V_g$. The curves are shifted by 1.25 μC/cm² for clarity. (b) Electric field dependence of $R_s$ for different maximum values of $V_g$ at 2 K. The curves are shifted by 150 Ω/sq for clarity.

A key aspect of STO-based 2DEGs is the strong electric field tunability with the voltage from e.g. a back-gate[30]. Remarkably, upon subjecting our samples to the electric field from a back-gate at $T$ = 2 K, we observe a very clear modulation of the sheet resistance with a visible hysteresis (see Figure 4(b)) as the electric field is cycled between different maximal values. At remanence after applying $E$ ±1.4 kV/cm, maximal values are found for the remanent sheet resistance and remanent polarization of $\Delta R_s$ = 72 Ω/sq and $2P_r$ = 0.85 µC/cm$^2$, respectively. Hall measurements at electrical remanence after applying ±1.4 V/cm reveal a relative carrier density modulation $\Delta n_s$=5.68 10$^{12}$ cm$^{-2}$, which is more than 50% of the value expected from the field effect when considering the remanent polarization of the Ca:STO (2$P_r$/e=1.075 10$^{13}$ cm$^{-2}$). The corresponding efficiency is among the highest reported in the literature[31–33].

It is tempting to refer to our 2DEG as a pristine ferroelectric 2DEG, as the 2DEG is formed within a ferroelectric material, Ca:STO, and because its transport properties show clear signatures of ferroelectricity (kink at $T_C$ in the $R$ vs $T$ data, hysteresis in the sheet resistance, etc). Yet, our data cannot unambiguously prove that the switchable polar state of stoichiometric Ca:STO survives in the 2DEG region where oxygen vacancies are present. We note however that an actual ferroelectric state has been suggested to coexist with metallicity in related compounds such as BaTiO$_{3-\delta}$ (from experiments[34,35] and theory[36]) and in bulk Sr$_{0.991}$Ca$_{0.009}$TiO$_{3-\delta}$[37]. In particular in BaTiO$_{3-\delta}$, theory indicates that ferroelectric displacements are sustained up to the critical concentration of 0.11 electron per unit cell[36]. Assuming a 2DEG thickness of 4 unit cells[12] our Hall data correspond to a carrier density of 0.04 electron per unit cell, below this critical value. It is thus possible that our Ca:STO based 2DEG corresponds to the first realization of a "ferroelectric" metal, as proposed by Anderson and Blount in 1965[38], and more recently for other systems[39–43], that is a material that exhibits cooperative polar displacements – which would produce a polarization in an insulator – and is at the same time metallic. Additional experiments using local probes of the polar displacements in the 2DEG region are required to confirm this possibility.

In summary, we have demonstrated the realization of a 2DEG based on ferroelectric Ca:STO, with gate-tunable transport properties and different resistance states at remanence. Our results bring a new degree of freedom to functionalize further STO 2DEGs and control the spin-charge interconversion properties[12,44] and the superconducting response[22,45] by ferroelectricity. Future studies may aim at calculating and characterizing the electronic structure of such ferroelectric 2DEGs, explore the role of Ca-substitution on the 2DEG properties, and possibly seek for room-temperature ferroelectricity, e.g. using strain[20,46].

**ACKNOWLEDGEMENTS**

This work received support from the ERC Consolidator grant no. 615759 "MINT", the ERC Advanced grant n° 833973 "FRESCO", the QUANTERA project "QUANTOX", the French Research Agency (ANR) as part of the "Investissement d'Avenir" program (LABEX NanoSaclay, ref ANR-10-LABX-0035) through project "AXION" and the Laboratoire d'Excellence LANEF (ANR-10-LABX-51-01) and ANR project OISO (ANR-17-CE24-0026-03). F. Trier acknowledges support by research grant VKR023371 (SPINOX) from VILLUM FONDEN.

**REFERENCES**

1. Hemberger, J., Lunkenheimer, P., Viana, R., Böhmer, R. & Loidl, A. Electric-field-dependent dielectric constant and nonlinear susceptibility in SrTiO$_3$. *Physical Review B* **52**, 13159–13162 (1995).

2. Müller, K. A. & Burkard, H. SrTiO$_3$ : An intrinsic quantum paraelectric below 4 K. *Phys. Rev. B* **19**, 3593–3602 (1979).

3. Bednorz, J. G. & Müller, K. A. Sr$_{1-x}$Ca$_x$TiO$_3$. An XY Quantum Ferroelectric with Transition to Randomness. *Physical Review Letters* **52**, 2289–2292 (1984).

4. Itoh, M. *et al.* Ferroelectricity Induced by Oxygen Isotope Exchange in Strontium Titanate Perovskite. *Physical Review Letters* **82**, 3540–3543 (1999).

5. Frederikse, H. P. R., Thurber, W. R. & Hosler, W. R. Electronic transport in strontium titanate. *Physical Review* **134**, A442–A445 (1964).

6. Tufte, O. N. & Chapman, P. W. Electron mobility in semiconducting strontium titanate. *Physical Review* **155**, 796–802 (1967).

7. Trier, F., Christensen, D. V. & Pryds, N. Electron mobility in oxide heterostructures. *J. Phys. D: Appl. Phys.* **51**, 293002 (2018).

8. Schooley, J. F., Hosler, W. R. & Cohen, M. L. Superconductivity in Semiconducting SrTiO$_3$. *Physical Review Letters* **12**, 474–475 (1964).

9. Lin, X., Zhu, Z., Fauqué, B. & Behnia, K. Fermi Surface of the Most Dilute Superconductor. *Physical Review X* **3**, (2013).

10. Ohtomo, A. & Hwang, H. Y. A high-mobility electron gas at the LaAlO$_3$/SrTiO$_3$ heterointerface. *Nature* **427**, 423–426 (2004).

11. Rödel, T. C. *et al.* Universal Fabrication of 2D Electron Systems in Functional Oxides. *Advanced Materials* **28**, 1976–1980 (2016).

12. Vaz, D. C. *et al.* Mapping spin–charge conversion to the band structure in a topological oxide two-dimensional electron gas. *Nat. Mater.* **18**, 1187-1193 (2019)

13. Santander-Syro, A. F. *et al.* Two-dimensional electron gas with universal subbands at the surface of $SrTiO_3$. *Nature* **469**, 189–193 (2011).

14. Yu, L. & Zunger, A. A polarity-induced defect mechanism for conductivity and magnetism at polar–nonpolar oxide interfaces. *Nature Communications* **5**, 5118 (2014).

15. Edge, J. M., Kedem, Y., Aschauer, U., Spaldin, N. A. & Balatsky, A. V. Quantum Critical Origin of the Superconducting Dome in $SrTiO_3$. *Phys. Rev. Lett.* **115**, 247002 (2015).

16. Tra, V. T. *et al.* Ferroelectric Control of the Conduction at the $LaAlO_3$/$SrTiO_3$ Heterointerface. *Advanced Materials* **25**, 3357–3364 (2013).

17. Kim, S.-I. *et al.* Non-Volatile Control of 2DEG Conductivity at Oxide Interfaces. *Advanced Materials* **25**, 4612–4617 (2013).

18. Wang, S. *et al.* Ferroelectric Polarization-Modulated Interfacial Fine Structures Involving Two-Dimensional Electron Gases in $Pb(Zr,Ti)O_3$/$LaAlO_3$/$SrTiO_3$ Heterostructures. *ACS Applied Materials & Interfaces* **10**, 1374–1382 (2018).

19. Russell, R. *et al.* Ferroelectric enhancement of superconductivity in compressively strained $SrTiO_3$ films. *Phys. Rev. Materials* **3**, 091401 (2019).

20. Pertsev, N. A., Tagantsev, A. K. & Setter, N. Phase transitions and strain-induced ferroelectricity in $SrTiO_3$ epitaxial thin films. *Physical Review B* **61**, R825–R829 (2000).

21. Verma, A., Raghavan, S., Stemmer, S. & Jena, D. Ferroelectric transition in compressively strained $SrTiO_3$ thin films. *Appl. Phys. Lett.* **107**, 192908 (2015).

22. Rischau, C. W. *et al.* A ferroelectric quantum phase transition inside the superconducting dome of $Sr_{1-x}Ca_xTiO_{3-\delta}$. *Nature Physics* **13**, 643–648 (2017).

23. Dunnett, K., Narayan, A., Spaldin, N. A. & Balatsky, A. V. Strain and ferroelectric soft-mode induced superconductivity in strontium titanate. *Phys. Rev. B* **97**, 144506 (2018).

24. Sing, M. *et al.* Profiling the Interface Electron Gas of $LaAlO_3/SrTiO_3$ Heterostructures with Hard X-Ray Photoelectron Spectroscopy. *Physical Review Letters* **102**, 176805 (2009).

25. Vaz, D. C. *et al.* Tuning Up or Down the Critical Thickness in $LaAlO_3/SrTiO_3$ through In Situ Deposition of Metal Overlayers. *Adv. Mater.* **29**, 1700486 (2017).

26. Bianchi, U., Kleemann, W. & Bednorz, J. G. Raman scattering of ferroelectric $Sr_{1-x}Ca_xTiO_3$, x=0.007. *J. Phys.: Condens. Matter* **6**, 1229–1238 (1994).

27. Biscaras, J. *et al.* Limit of the electrostatic doping in two-dimensional electron gases of $LaXO_3$(X = Al, Ti)/$SrTiO_3$. *Scientific Reports* **4**, 6788 (2015).

28. Basletic, M. *et al.* Mapping the spatial distribution of charge carriers in $LaAlO_3/SrTiO_3$ heterostructures. *Nature Mater* **7**, 621–625 (2008).

29. Ahadi, K. *et al.* Enhancing superconductivity in $SrTiO_3$ films with strain. *Sci. Adv.* **5**, eaaw0120 (2019).

30. Caviglia, A. D. *et al.* Electric field control of the $LaAlO_3/SrTiO_3$ interface ground state. *Nature* **456**, 624–627 (2008).

31. Crassous, A. *et al.* Nanoscale Electrostatic Manipulation of Magnetic Flux Quanta in Ferroelectric/Superconductor $BiFeO_3/YBa_2Cu_3O_{7-\delta}$ Heterostructures. *Phys. Rev. Lett.* **107**, 247002 (2011).

32. Yamada, H. *et al.* Ferroelectric control of a Mott insulator. *Sci Rep* **3**, 2834 (2013).

33. Hong, X. Emerging ferroelectric transistors with nanoscale channel materials: the possibilities, the limitations. *Journal of Physics: Condensed Matter* **28**, 103003 (2016).

34. Kolodiazhnyi, T. Insulator-metal transition and anomalous sign reversal of the dominant charge carriers in perovskite $BaTiO_{3-\delta}$. *Physical Review B* **78**, 045107 (2008).

35. Kolodiazhnyi, T., Tachibana, M., Kawaji, H., Hwang, J. & Takayama-Muromachi, E. Persistence of Ferroelectricity in BaTiO$_3$ through the Insulator-Metal Transition. *Physical Review Letters* **104**, (2010).

36. Wang, Y., Liu, X., Burton, J. D., Jaswal, S. S. & Tsymbal, E. Y. Ferroelectric Instability Under Screened Coulomb Interactions. *Phys. Rev. Lett.* **109**, 247601 (2012).

37. Engelmayer, J. *et al.* Ferroelectric order versus metallicity in Sr$_{1-x}$Ca$_x$TiO$_{3-\delta}$ (x = 0.009). *Phys. Rev. B* **100**, 195121 (2019).

38. Anderson, P. W. & Blount, E. I. Symmetry Considerations on Martensitic Transformations: 'Ferroelectric' Metals? *Phys. Rev. Lett.* **14**, 217–219 (1965).

39. Shi, Y. *et al.* A ferroelectric-like structural transition in a metal. *Nature Materials* **12**, 1024–1027 (2013).

40. Keppens, V. 'Ferroelectricity' in a metal: Structural transitions. *Nature Materials* **12**, 952–953 (2013).

41. Benedek, N. A. & Birol, T. 'Ferroelectric' metals reexamined: fundamental mechanisms and design considerations for new materials. *J. Mater. Chem. C* **4**, 4000–4015 (2016).

42. Filippetti, A., Fiorentini, V., Ricci, F., Delugas, P. & Íñiguez, J. Prediction of a native ferroelectric metal. *Nat. Commun* **7**, 11211 (2016).

43. Fujioka, J. *et al.* Ferroelectric-like metallic state in electron doped BaTiO$_3$. *Sci. Rep.* **5**, 13207 (2015).

44. Lesne, E. *et al.* Highly efficient and tunable spin-to-charge conversion through Rashba coupling at oxide interfaces. *Nature Mater* **15**, 1261–1266 (2016).

45. Reyren, N. *et al.* Superconducting Interfaces Between Insulating Oxides. *Science* **317**, 1196–1199 (2007).

46. Haeni, J. H. *et al.* Room-temperature ferroelectricity in strained SrTiO$_3$. *Nature* **430**, 758–761 (2004).